\documentclass[aps,prl,reprint,superscriptaddress,floatfix,longbibliography]{revtex4-2}

\usepackage{bm}
\usepackage{amsmath}
\usepackage{graphicx}
\usepackage{siunitx}
\usepackage{url}
\usepackage{hyperref}
\usepackage{etoolbox}
\apptocmd{\sloppy}{\hbadness 10000\relax}{}{}

\graphicspath{{figures/}}
\begin{document}

\newcommand{\nanometer}{\,\si{\nano\meter}}
\newcommand{\micrometer}{\,\si{\micro\meter}}
\newcommand{\millimeter}{\,\si{\milli\meter}}
\newcommand{\gigahertz}{\,\si{\giga\hertz}}
\newcommand{\kilohertz}{\,\si{\kilo\hertz}}

\title{Long range magnetic dipole-dipole interaction mediated by a superconductor}
\author{Yoav Romach\normalfont\textsuperscript{\dag}}
\email[Corresponding author: ]{yoav.romach@mail.huji.ac.il}
\affiliation{The Racah Institute of Physics, The Hebrew University of Jerusalem, Jerusalem 9190401, Israel}
\author{Tal Wasserman}
\thanks{These authors contributed equally to this work.}
\affiliation{The Racah Institute of Physics, The Hebrew University of Jerusalem, Jerusalem 9190401, Israel}
\author{Shai Tishby}
\thanks{These authors contributed equally to this work.}
\affiliation{The Racah Institute of Physics, The Hebrew University of Jerusalem, Jerusalem 9190401, Israel}
\author{Nir Bar-Gill}
\affiliation{The Racah Institute of Physics, The Hebrew University of Jerusalem, Jerusalem 9190401, Israel}
\affiliation{Department of Applied Physics, Rachel and Selim School of Engineering, The Hebrew University of Jerusalem, Jerusalem 9190401, Israel}
\begin{abstract}
Quantum computation and simulation requires strong coherent coupling between qubits, which may be spatially separated. Achieving this coupling for solid-state based spin qubits is a long-standing challenge. Here we theoretically investigate a method for achieving such coupling, based on superconducting nano-structures designed to channel the magnetic flux created by the qubits. We detail semi-classical analytical calculations and simulations of the magnetic field created by a magnetic dipole, depicting the spin qubit, positioned directly below nanofabricated apertures in a superconducting layer. We show that such structures could channel the magnetic flux, enhancing the dipole-dipole interaction between spin qubits and changing its scaling with distance, thus potentially paving the way for controllably engineering an interacting spin system.
\end{abstract}
\maketitle

\section{Introduction}
Solid-state qubits have emerged as a potential quantum information processing architecture, with leading candidates such as atomic defects in bulk materials \cite{wrachtrup2006} and quantum dots \cite{Awschalom_qdots_2013}. High fidelity quantum control of such individual spin qubits has been demonstrated in various systems \cite{semiconductor_review_2020, awschalom_quantum_2018}. However, scalable coherent coupling of these qubits, required in order to produce robust two-qubit gates, still poses a significant challenge.
The direct magnetic coupling between spin qubits via the dipole-dipole interaction is relevant only for spins that are quite close (at the scale of 10 nm), as this interaction usually decays with the distance cubed. This has been demonstrated, e.g., for both Nitrogen-Vacancy (NV) centers in diamond \cite{neumann_Quantum_register_2010, dolde_room_temperature_2013} and for quantum dots \cite{Shulman_entanglement_qdots_2012, veldhorst_two_qubit_2015}.
Applications in quantum computing will require a high number of coupled qubits which can be easily addressed. This would be much easier to implement if the qubits are spatially separated to much longer distances than usually possible with dipole-dipole interaction. One such method is to use another system, such as photons, as a quantum bus \cite{Awschalom_qdots_2013, bernien2013, su_arbitrary_2018}. Another method is to use an external cavity or a ``floating gate'' \cite{dolde_floating_gates_2013, warren_long_distance_2019}.

\begin{figure}[htb]
	\begin{center}
		\includegraphics[width=1 \columnwidth]{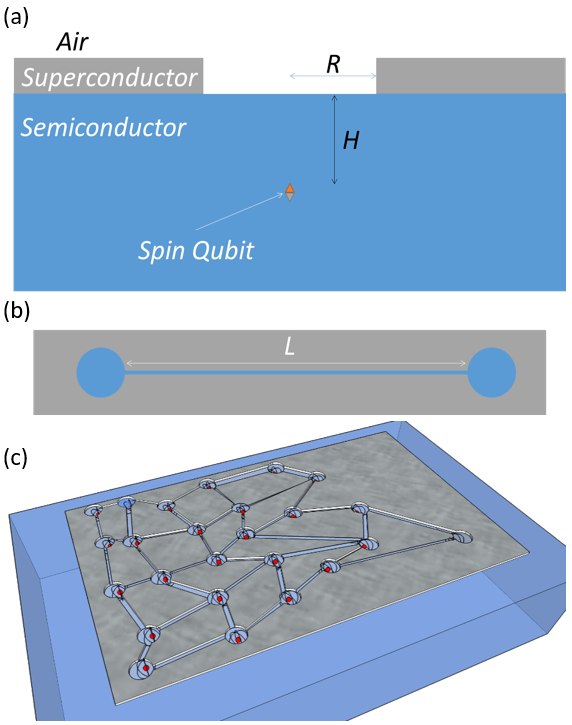}
		\protect\caption{(a) A side-way view of an spin qubit in a semiconductor underneath an aperture in a superconducting layer. Relevant length scales are noted with $R\sim H$. (b) A top-down view of a ``dog-bone'' structure composed of two apertures at a distance of $L$ and a thin channel that connects them. It is assumed that $L\gg R$. (c) An artist view of a possible large-scale implementation of a quantum device with many spin qubits coupled using such structures.}
		\label{fig:qubit_dogbone}
	\end{center}
\end{figure}

In this work we propose a method to increase the dipole-dipole interaction to much longer distances. We propose to place dipoles inside fabricated nano-structures in a thin layer of a superconductor. The superconductor would guide the field lines created by the dipoles into the nano-structures, a phenomena known as flux focusing, and increase the interaction between the dipoles.
A possible structure would be a ``dog-bone'' shaped structure shown in Fig.\,\ref{fig:qubit_dogbone}. Fig.\,\ref{fig:qubit_dogbone}(a) is a side-way view of ``dog-bone'' structure fabricated directly above an NV center or another bulk-semiconductors based qubit. For quantum dots or other dipoles, this device could be fabricated on a Si wafers with the quantum dots later placed inside the apertures. Fig.\,\ref{fig:qubit_dogbone}(b) shows a top-down view of a ``dog-bone'' structure.
This structure, with a ferromagnet instead of a superconductor, has already been proposed as a possible way to create long-range coherent interaction between NV centers \cite{dogbone_ferromagnet_2013}. It has also been proposed as a long-distance coupler between phosphorous based qubits in Si \cite{semiconductor_review_2020}.
We also note that this device could be used for sensing application, with the sensor placed in one aperture and the sample placed in another aperture. This could be useful whenever the sample requires a ``clean'' area far from the sensor. In the case of NV centers, such a realistic scenario could arise for a sample which might be damaged by the laser used to address the NV center.

\begin{figure}[htb]
	\begin{center}
		\includegraphics[width=1 \columnwidth]{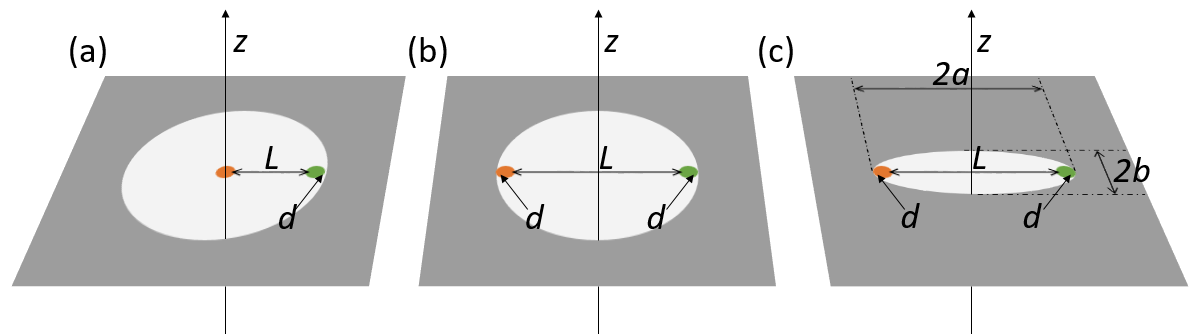}
		\protect\caption{The three cases addressed in this work. The dipole creating the field is marked as an orange dot and the point at which the field is measured is marked as a green dot. This point is at a distance $d$ from the right edge of the aperture in all three cases. (a) A round aperture with the dipole in the center. (b) A round aperture with the dipole at a distance $d$ from the left edge. (c) An elliptical aperture with the dipole at a distance $d$ from the left edge.
		Relevant distances are noted.}
		\label{fig:Schematics}
	\end{center}
\end{figure}

We look at three simpler scenarios illustrated in Fig.\,\ref{fig:Schematics}: A round aperture with a dipole in the center, a round aperture with a dipole next to the left edge, and an elliptical aperture with a dipole next to the left edge. In all three cases the 2nd dipole would be placed next to the right edge of the aperture. We note that the difference between the ``dog-bone'' behavior and the ellipse behavior should be small when the ellipse eccentricity is close to 1, due to the similar one-dimensional confinement of the magnetic field. Having a circular aperture at the edge of the ellipse (a ``dog-bone'' structure) would ease the fabrication and the accurate placement of the structure relative to the qubit, but should not change the power law scaling of the coupling.
We show that by placing the dipoles inside these fabricated nano-structures the dipole-dipole interaction is significantly enhanced.

Calculating the currents and magnetic field in the vicinity of a superconductor is a challenging task. Analytical solutions are generally only available for simple symmetric geometries, such as planes, spheres and cylinders. Those cases have been widely studied, mostly for calculating the interaction and levitation forces \cite{Examples1,Examples2,Examples3,Examples4,Examples5,Examples6,Examples7,Examples8,Examples9,Examples10,Examples11,Examples12}. Moreover, numerical solutions are usually either for bulk samples \cite{caputo_screening_2013} or simple geometries \cite{xu_magnetic_2008, tomkow_numerical_2019, wang_stress_2020}, and are generally very time and/or resource intensive. 

Here we present a method for calculating the exact analytical solution for the magnetic field arising from a single dipole inside a circular aperture in a superconducting thin film, under the assumption of zero penetration depth. We solve for a dipole at the center of the aperture and then at a shifted position, and show that the flux focusing and field confinement can be used to dramatically increase the dipole-dipole interaction. In addition, we numerically solve the problem with a finite penetration depth and show a very good agreement with the analytic calculations. Finally, we extend the numerical results to an elliptical aperture and show an even further increase in the interaction strength. 

During the preparation of this article we became aware of a recent similar work used to find the flux focusing in a parallel SQUID array\cite{muller_theoretical_2020}.

\section{Analytical Analysis}
A known form of the London Equations, which describes the relation between the vector potential $\textbf{A}$ and the current density $\textbf{J}$ in a superconductor is \cite{london_electromagnetic_1935}
\begin{equation}
\boldsymbol{A}=-\mu_{0}\lambda^{2}\boldsymbol{J},
\label{eq:London}
\end{equation}
where $\mu_0$ is the vacuum permeability and $\lambda$ is the superconductor penetration depth. Thus $\lambda\rightarrow 0 \Rightarrow \boldsymbol{A}|_{\sigma} \rightarrow 0$, where $\sigma$ denotes the boundary of the superconductor,
implying that the limit of zero penetration depth of a superconductor results in a vanishing vector potential on the boundaries of the superconductor \footnote{One needs to be a bit cautious when writing this form of the London equations for the geometry of a superconductor that is not simply connected, since in that case it is correct only if for every closed path $\Gamma$ around an aperture, $\boldsymbol{A}$ still satisfies $\oint_{\Gamma}\boldsymbol{A}\cdot dl=\phi_{\Gamma}$, where $\phi_{\Gamma}$ is the flux enclosed by the path \cite{LondonTheory}. This requirement is necessary since far from the aperture, $\boldsymbol{J}\rightarrow0$, so plugging in the London equation in the integral gives zero, instead of $\phi_{\Gamma}$. In our specific case of a single aperture and an infinite film, the flux is zero anyway and so this London equation does hold}.

Under these conditions, we can base our calculation off of the method of the Dirichlet electrostatic Green function \cite{Zangwill}. In the Coulomb Gauge we have $\nabla^{2}\boldsymbol{A}(\boldsymbol{r})=-\mu_{0}\boldsymbol{J}(\boldsymbol{r})$, where each of the Cartesian components of the vector potential behaves like an electrostatic potential with $\frac{\rho(\boldsymbol{r})}{\epsilon_{0}}\rightarrow\mu_{0}J_{i}(\boldsymbol{r})$ ($\epsilon_0$ is the vacuum permittivity). Using this method, when given a current density $\boldsymbol{J}(\boldsymbol{r})$ and a Green function $G(\boldsymbol{r},\boldsymbol{r}\boldsymbol{'})$, one is able to calculate the vector potential and the resulting magnetic field:
\begin{equation}
\boldsymbol{B}(\boldsymbol{r})=\nabla\times\boldsymbol{A}(\boldsymbol{r})=\nabla\times\mu_{0}\epsilon_{0}\int_{V}d^{3}r'\boldsymbol{J}(\boldsymbol{r}\boldsymbol{'})G(\boldsymbol{r},\boldsymbol{r}\boldsymbol{'}).
\label{eq:B_from_int_JG}
\end{equation}
In our case, we have a superconducting film lying in the $xy$ plane, with a circular aperture of radius $R$, i.e.: $\sigma=\{z=0\ \cap\ \rho\geq R\}$. The current density of a point magnetic dipole $\boldsymbol{m}$, located at $\boldsymbol{r_0}$, is given by \cite{Zangwill} $\boldsymbol{J}(\boldsymbol{r})\boldsymbol{=}-\boldsymbol{m}\times\nabla\delta(\boldsymbol{r}-\boldsymbol{r_0})$.

The problem of finding the Green Function of a semi-infinite film, $\{z=0\ \cap\ x\geq 0\}$, is treated in \cite{Semi_Infinite}. Following the process in \cite{Green_Function}, one is able to apply the Kelvin inversion transformation that can be used to generate new solutions for the Poisson equation from known ones \cite{Inversion}. This transformation inverts the space relative to a predefined sphere. By applying the Kelvin inversion we are able to transform the Green function of the semi-infinite film to one of an infinite film with a circular aperture, i.e for $\sigma$. For more details, see appendix.

Attempting to generalize this approach in order to solve an asymmetric aperture (more specifically an elliptic aperture) is not a trivial task. Naively stretching one of the $x$ or $y$ coordinates in the circular Green function generates a new function which vanishes outside of an elliptical aperture, but no longer satisfies the Poisson equation. There is a modified Kelvin inversion transformation, which inverts the space relative to an ellipsoid rather than a sphere, but unfortunately, this transformation is non-conformal \cite{Elliptic_Inversion}. Non-conformality implies that it will not generate a correct Poisson equation solution in the new elliptical geometry. There are many known 2D conformal transformations that are able to transform a circular aperture to an asymmetric aperture, but  applying them to the $x,y$ coordinates of a 3D Green function will not be conformal. To the best of our knowledge there is no 3D conformal transformation which can transform a circle into an ellipse. We therefore proceed with the analysis of a circular aperture, with a centered and off-center dipole, and then continue with numerical calculations for the elliptic case.

\subsection{Dipole at the center of an aperture}
The simplest case to solve is the case illustrated in Fig.\,\ref{fig:Schematics}(a): a dipole in the center of a round aperture.
Plugging in the Green function from \cite{Green_Function} and the current density of a dipole \cite{Zangwill} located at the origin into Eq.\,\ref{eq:B_from_int_JG}, we get (For a detailed derivation, see appendix):
\begin{equation}
\boldsymbol{A}(\boldsymbol{r})=\mu_{0}\epsilon_{0}\boldsymbol{m}\times\nabla_{\boldsymbol{r'}}G(\boldsymbol{r},\boldsymbol{r}\boldsymbol{'})|_{\boldsymbol{r}\boldsymbol{'}=\boldsymbol{0}}=\frac{\mu_{0}}{4\pi}\frac{\boldsymbol{m}\times\boldsymbol{\hat{n}}(\boldsymbol{r})}{r^{2}}.
\label{eq:A_for_centered_dipole}
\end{equation}
Here,
\begin{align*}
\boldsymbol{n}(\boldsymbol{r}) & \equiv C(\boldsymbol{r)}\rho\hat{\rho}+z\hat{z} \\
\\
C(\boldsymbol{r)} & \equiv \frac{2}{\pi}\biggl(\textbf{tan}^{-1}\Bigl(\alpha(\boldsymbol{r})\Bigr)+\frac{\alpha(\boldsymbol{r})}{1+\alpha(\boldsymbol{r})^{2}}\biggr)
\\
\alpha(\boldsymbol{r}) & \equiv \\ 
& \frac{1}{\sqrt{2}r}\sqrt{R^{2}-r^{2}+\sqrt{\bigl[z^{2}+\bigl(\rho+R\bigr)^{2}\bigr]\bigl[z^{2}+\bigl(\rho-R\bigr)^{2}\bigr]}}
\end{align*}

\begin{figure}[htb]
	\begin{center}
		\includegraphics[width=1 \columnwidth]{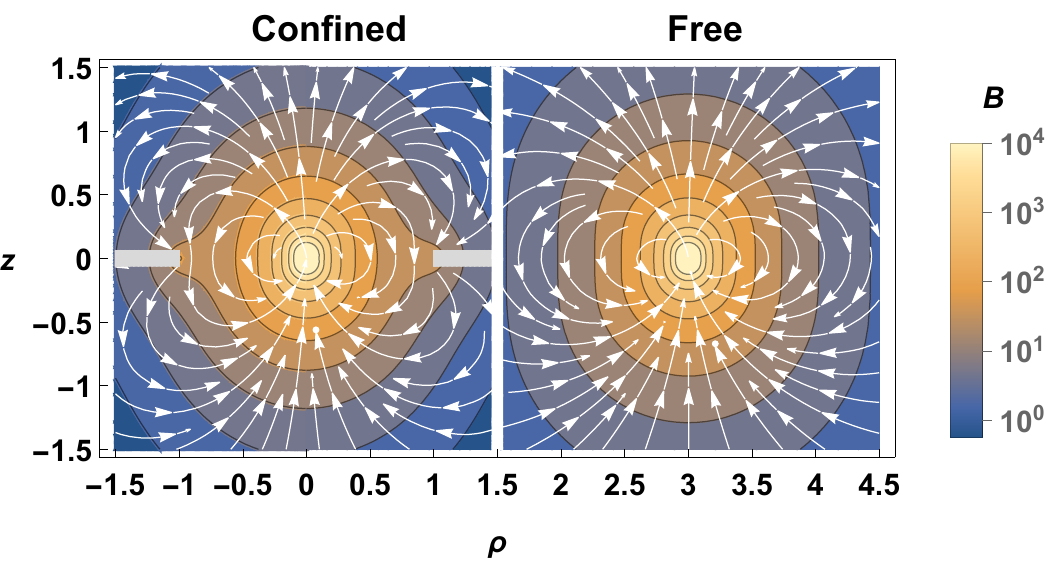}
		\protect\caption{The analytical solution of the magnetic dipole field in a circular aperture of a superconducting film, calculated from Eq.\,\ref{eq:B_for_centered_dipole}, for $\mu_{0}m=100, R=1$. The dipole is pointing towards $\hat{z}$. The stream plot provides the direction of the field, and the contour plot is scaled relative to the magnitude of the field. We see the comparison to the free magnetic dipole field in the right side plot. It is evident visually that the superconductor confines the magnetic flux.}
		\label{fig:analytical_streamplot_for_round_hole}
	\end{center}
\end{figure}

It is instructive to introduce the vector $\boldsymbol{\hat{n}}(\boldsymbol{r})$, so one can easily notice that the vector potential In Eq.\,\ref{eq:A_for_centered_dipole} has the same form as a free magnetic dipole vector potential, with $\boldsymbol{\hat{n}}(\boldsymbol{r})\rightarrow\boldsymbol{\hat{r}}$. The entire effect of the superconductor is reflected in the coefficient $C(\boldsymbol{r)}$ in the $\hat{\rho}$ direction, which is a location dependent scaling factor. We see that on the boundary $\sigma$, $\alpha=0$, so $\boldsymbol{n}|_{\sigma}=0$, and then $\boldsymbol{A}|_{\sigma}=0$, consistent with our construction of $G(\boldsymbol{r},\boldsymbol{r}\boldsymbol{'})$. As a sanity check, the limit for an infinitely large aperture $R\rightarrow\infty$ (or equivalently, $r\rightarrow0$)
\[
\implies\alpha\rightarrow\infty\implies C(\boldsymbol{r)}\rightarrow1\implies\boldsymbol{\hat{n}}(\boldsymbol{r})\rightarrow\boldsymbol{\hat{r}}
\]
reproduces the expression of a free dipole.
Now for the magnetic field, taking the curl of Eq.\,\ref{eq:A_for_centered_dipole} we obtain
\begin{equation}
\begin{split}
\boldsymbol{B}(\boldsymbol{r})=\frac{\mu_{0}}{4\pi r^{3}} \biggl[ 3\bigl(\boldsymbol{m}\cdot\boldsymbol{\hat{r}}\bigr)\hat{\boldsymbol{n}}(\boldsymbol{r}) - \bigl(\boldsymbol{m}\cdot\nabla\bigr)\boldsymbol{n}(\boldsymbol{r}) \\
+ \boldsymbol{m} \Bigl(\nabla\cdot\boldsymbol{n}(\boldsymbol{r}) - 3\bigl(\hat{\boldsymbol{r}}\cdot\boldsymbol{\hat{n}}(\boldsymbol{r})\bigr)\Bigr) \biggr].
\end{split}
\label{eq:B_for_centered_dipole}
\end{equation}
When considering the limit of an infinitely large aperture, $\boldsymbol{\hat{n}}(\boldsymbol{r})\rightarrow\boldsymbol{\hat{r}}$ as before, the two right terms cancel each other out and $\bigl(\boldsymbol{m}\cdot\nabla\bigr)\boldsymbol{n}(\boldsymbol{r})\rightarrow\boldsymbol{m}$,
resulting in the known expression for the magnetic dipole field.
In Fig.\,\ref{fig:analytical_streamplot_for_round_hole} we plot the magnetic field calculated from Eq.\,\ref{eq:B_for_centered_dipole}, comparing the case of a small aperture (normalized size $R=1$) to the free dipole case (without an aperture).

\begin{figure}[htb]
	\begin{center}
		\includegraphics[width=1 \columnwidth]{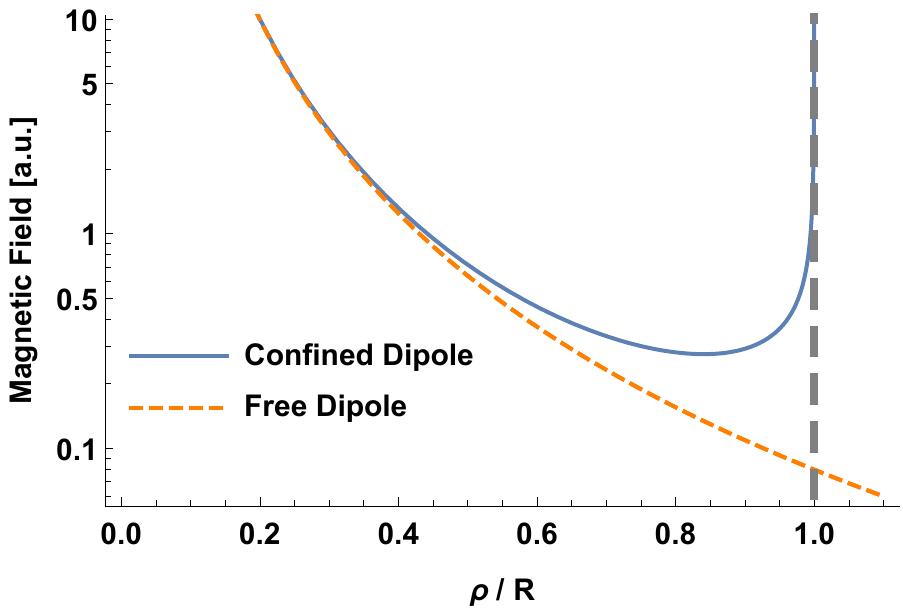}
		\protect\caption{The analytical solution of the magnetic dipole field in a circular aperture of a superconducting film, calculated from Eq.\,\ref{eq:Bmz_for_centered_dipole}. The dashed vertical line indicates the edge of the superconductor. The field vanishes on the superconductor, since no field lines can cross it. Close to the center, for which the limit of $\rho\rightarrow0$ (or $R\rightarrow\infty$) is valid , we find a behavior consistent with a free dipole, approaching $-\frac{\mu_{0}m}{4\pi\rho^{3}}$. Closer to the edge we see that the field is approaching infinity as can be seen from Eq.\,\ref{eq:1/sqrtd}} \label{fig:analytical_Bmz_for_round_hole}
	\end{center}
\end{figure}

Restricting ourselves to the $xy$ plane, we can derive a relatively simple analytical expression for the magnetic field inside the aperture. Due to the superconductor fully blocking the field in the $\hat{z}$ direction, the strongest confinement occurs for a dipole oriented along the $\hat{z}$ direction (See appendix for the full derivation). Starting from Eq.\,\ref{eq:B_for_centered_dipole} and setting the dipole to point along the $\hat{z}$ direction, we derive the magnetic field in the $xy$ plane as
\begin{equation}
{B}_{z}(\rho,\phi,0)=-\frac{\mu_{0}m}{2\pi^{2}\rho^{3}}\Biggl[\textbf{cos}^{-1}\bigl(\frac{\rho}{R}\bigr)+\frac{\frac{\rho}{R}\bigl(1+\bigl(\frac{\rho}{R}\bigr)^{2}\bigr)}{\sqrt{1-\bigl(\frac{\rho}{R}\bigr)^{2}}}\Biggr].
\label{eq:Bmz_for_centered_dipole}
\end{equation}
This equation is plotted in Fig.\,\ref{fig:analytical_Bmz_for_round_hole}, stressing the flux focusing and thus enhanced magnetic field near the edge of the aperture.

We can now use Eq.\,\ref{eq:Bmz_for_centered_dipole} to infer the strength of the field confinement near the edge, if we maintain a constant distance from the edge, denoted by $d$, and increase the aperture’s radius:
\begin{equation}
{B}_{z}(\rho\rightarrow{R}\rightarrow\infty,0,0)=-\frac{\mu_{0}m}{\sqrt{2}\pi^{2}}\frac{1}{\sqrt{d}R^{\frac{5}{2}}}+O\left(\frac{1}{R^{\frac{7}{2}}}\right).
\label{eq:1/sqrtd}
\end{equation}
We find that the power-law scaling of the magnetic field with the radius has increased from the usual dipole power-law of $-3$ to $-2.5$. This result is plotted in Fig.(\ref{fig:analytical_powerlaw_decay}).

\subsection{Dipole at the side of an aperture}
\begin{figure}[htb]
	\begin{center}
		\includegraphics[width=1 \columnwidth]{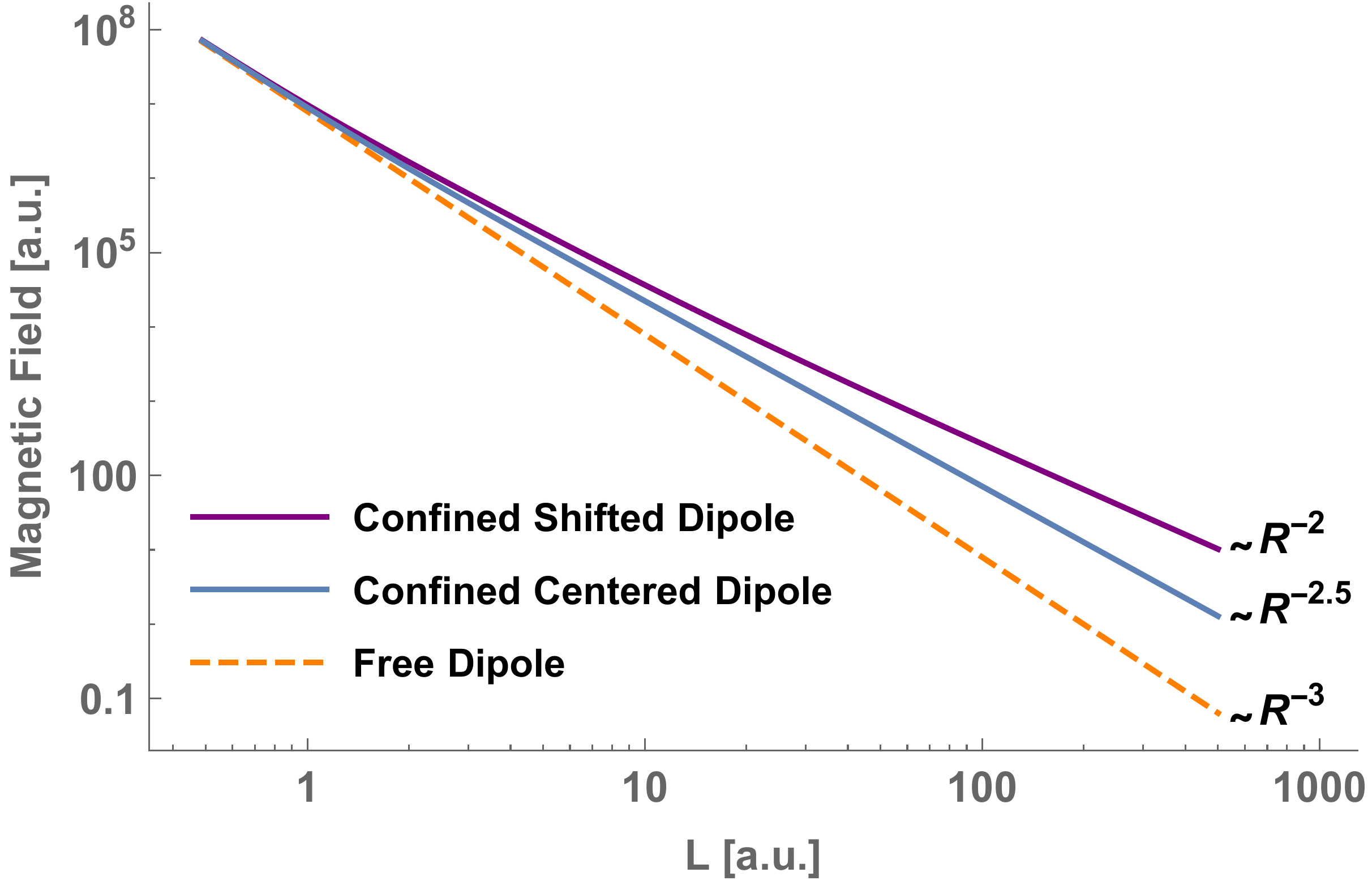}
		\protect\caption{The magnetic field as a function of the distance $L$ between the dipole and the measured point plotted for three cases: A free dipole, a dipole at the center of the aperture, and a dipole next to aperture's edge. The superconductor aperture's radius $R$  varies and the distance of the measured point from the edge of the aperture is constant $d=1$. For the centered dipole, $R = L + d$. For the shifted dipole, $R = \frac{L}{2} + d$ where $d$ is the distance of both the dipole and the measured point from the aperture's edges.
		When the radius is large enough we observe the improved power-law scaling and the magnetic field enhancement indicated in Eq.\,\ref{eq:1/sqrtd}-\ref{eq:1/d}.} \label{fig:analytical_powerlaw_decay}
	\end{center}
\end{figure}

Shifting to a dipole which is off-center, e.g. at the left edge of a round aperture [Fig.\,\ref{fig:Schematics}(b)] is relatively simple.
Starting with Eq.\,\ref{eq:B_from_int_JG}, we use a dipole current density which is shifted in the $\hat{x}$ direction by $x_0$. The derivation is similar to that of the non-shifted case but the solutions for the vector potential and the field, calculated with Mathematica \cite{Mathematica12p1}, are given as complicated expressions which are too long to include fully. Nevertheless, we verify that by substituting $x_0 = 0$ we again obtain the results of a centered dipole (Eq.\,\ref{eq:A_for_centered_dipole}-\ref{eq:Bmz_for_centered_dipole}). Moreover, the magnetic field can be plotted and simplified next to the edge: we derive the expression for a dipole shifted to $x_0 = d-R$ (a distance $d$ from the left edge of the aperture) and we look at the magnetic field symmetrically near the other edge, at the point $x = R-d$, where $d$ is assumed small relative to the radius $R$. Using Mathematica \cite{Mathematica12p1} we obtain a simplified series expansion
\begin{equation}
{B}_{z}(x\rightarrow{R}\rightarrow\infty,0,0)=-\frac{\mu_{0}m}{4 \pi^2}\frac{1}{d R^2}+O\left(\frac{1}{R^{3}}\right).
\label{eq:1/d}
\end{equation}
We find a significant enhancement of the magnetic field, with the power-law scaling improving from the usual dipole power-law of $-3$ to $-2$. This change in scaling is depicted in Fig.\,\ref{fig:analytical_powerlaw_decay}.

In order to gain insight into this behavior, we can start by examining a dipole which is shifted to $x_0 = R-d$, and calculate the magnetic field at the origin. We can see that this problem is symmetric to the problem solved in the previous section, by switching between the dipole and the point at which we measure the field. Therefore, in this case we get the exact same behaviour as in Eq.\,\ref{eq:1/sqrtd}.
We can now combine both effects: shifting the dipole to distance $d$ from one edge, and calculating the field at a distance $d$ from the second edge, leading to a $\sqrt{R}$ improvement from each effect, resulting in Eq.\,\ref{eq:1/d}.

\section{Numerical Simulations}
We augment our analytical analysis with numerical simulations, to address an optimized aperture geometry which is not circular (but rather elliptical) and thus cannot be addressed analytically.
The simulations were performed by solving the London equations as described in \cite{brandt_2005}. This method is valid for a magnetic field much smaller than the upper critical field $H_{c2}$ of the superconductor, and for a superconductor much larger than its coherence length $\xi$.

The physical quantity obtained from the simulation is the thin film current $\boldsymbol{J}(x,y) = \int dz \boldsymbol{j}(x,y,z) = (J_x, J_y)$. If the thickness is nearly constant with thickness $\Delta$, and the film is thin enough such that $\boldsymbol{j}(x,y,z)$  is not dependent on $z$, we can approximate $\boldsymbol{J}(x,y) = \boldsymbol{j}(x,y,z) \Delta$. Since the divergence of the current is zero $\nabla \cdot \boldsymbol{J} = 0$, we can express it in terms of a scalar potential called the stream function $g(x,y)$
\begin{equation}
\boldsymbol{J} = -\hat{z} \times \nabla g = \nabla \times (\hat{z} g) = (\partial g / \partial y, -\partial g/ \partial x).
\label{eq:stream_function}
\end{equation}

The simulations find the current stream function $g(x,y)$ and the effective magnetic field $H_z (x,y)$ resulting from an arbitrarily shaped superconducting film with an applied magnetic field $H_a (x,y)$. The simulations are carried out on a non-equidistant grid and use a matrix inversion method with matrix size $(N_x \times N_y) \times (N_x \times N_y)$, where $N_{x,y}$ are the numbers of grid points in directions $(x,y)$. This process is very resource intensive and therefore only works for small grids (usually grid sizes of $\lesssim 100 \times 100$ points).

The simulation invert a matrix whose non-diagonal terms are multiplied by the effective penetration depth (also known as the 2D screening length or Pearl length) $\Lambda = \frac{\lambda^2}{\Delta}$ \cite{pearl_current_1964}, where $\Delta$ is again the superconducting film thickness and $\lambda$ is the London penetration depth. Therefore, in order for the matrix inversion to be smooth, we require $\Lambda \gg 0$. The London penetration depth $\lambda$ can be smaller than the film thickness $\Delta$ as long as it is of the same order of magnitude.

For our simulations, we choose $\Delta=80 \nanometer$ and $\lambda = 50 \nanometer$, which correspond to the London penetration depth for a clean Niobium layer of the given thickness.
The superconducting film is taken to be $90$ times the radius of the aperture and the entire simulation grid size is taken to be $100$ times the radius. The applied magnetic field is taken to be the field created from a magnetic dipole pointing in the $\hat{z}$ direction: $H_a(x,y) = \frac{m}{2\pi r^3} \hat{z}$. The magnetization is of a single spin 1 particle: $m = 2 g \mu_B$ where $g$ is the electron g-factor and $\mu_B$ is the Bohr magneton.

In the symmetric case, for which the dipole is at the center of a round aperture, there are $100$ grid points per axis. The points are chosen such that there is a higher density of points next to the edges, where the gradient is higher. For the other cases, the number of points per axis is not the same. The simulations were executed with varying numbers of grid points to ensure stability and convergence (For a detailed explanation of the simulation and the choice of grid points see \cite{brandt_2005}).

\subsection{Comparison between numerical simulations and analytical calculation}
\begin{figure}[htb]
	\begin{center}
		\includegraphics[width=1 \columnwidth]{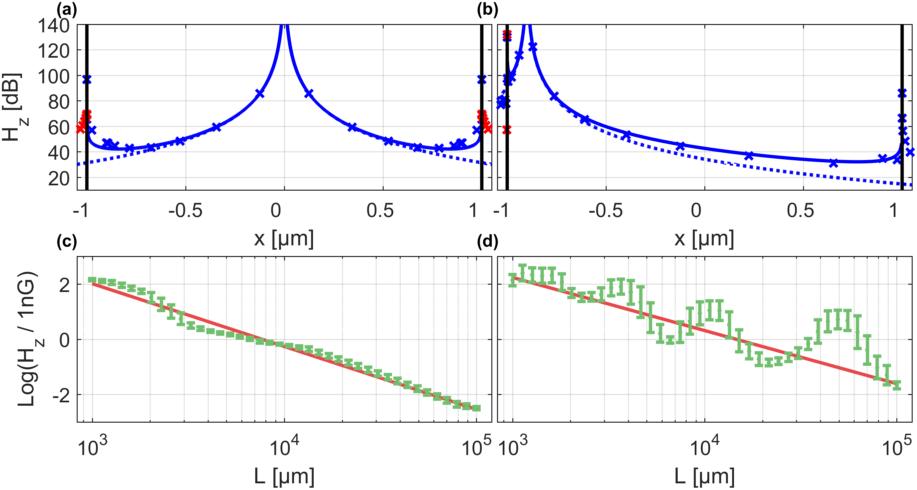}
		\protect\caption{(a,b) The effective magnetic field $H_z$ inside a round aperture with a radius of $1000 \nanometer$ created by a dipole (a) in the center of the aperture and (b) at a distance of $d=100 \nanometer$ from the left edge. The field is shown at the line $y = 5 \nanometer$. The y axis is the magnetic field, in dB, relative to 1 Gauss. The black lines indicate the edges of the aperture. The dashed curve shows the normal dipole decay and the solid curve depicts the results from the analytical analysis. The points are the results of the simulation. A blue (red) point indicates that the field is in the negative (positive) $\hat{z}$ direction.
		(c,d) The simulated magnetic field $H_z$ at a distance of $d$ from the right edge of the aperture as a function of $L$ - the distance between the dipole and the measured point. (c) A dipole at the center of the aperture. The radius varies as $R = L+d$. The line is fitted to the log of the data and the slope is $-2.3 \pm 0.1$. (d) A dipole at a distance of $d$ from the left edge of the aperture. The radius varies as $R = \frac{L}{2}+d$. The line is fitted to the log of the data and the slope is $-1.9 \pm 0.1$.} \label{fig:simulation_analytic}
	\end{center}
\end{figure}

We first simulated the case of a round aperture [Fig.\,\ref{fig:Schematics}(a,b)] in order to compare the numerical results with the results of the analytical analysis. As can be seen in Fig.\,\ref{fig:simulation_analytic}(a,b), the simulations and analytical results exhibit the same trend, with small deviations. 
The London penetration depth $\lambda$ has an effect of smearing the currents: a smaller value allows the currents to ``change'' on a shorter length-scale. This results in having the effective magnetic field more localized toward the edge of the aperture. This can partly explain the small discrepancy between the analytical results (which require $\lambda=0$) and the numerical method which require a $\lambda \gg 0$.

Next, the simulations were carried out on circular apertures with varying radii, with the dipole either located at the center or at a distance of $d=100 \nanometer$ from the left edge of the aperture. The magnetic field at a constant distance of $d$ from the right edge of the aperture was extracted from the data. The resulting data can be seen in Fig.\,\ref{fig:simulation_analytic}(c,d). The simulation results deviate from the analytics in certain cases, for which these results indicate a higher magnetic field inside the aperture. The cause of this discrepancy requires further study which is beyond the scope of this work. Nevertheless, we smooth the results by averaging the data using a moving window method, and the standard error was extracted using a moving std window.

We then fit the data using a weighted least squares regression and obtain a consistent power-law scaling to the one predicated by the analytical analysis (Eq.\,\ref{eq:1/sqrtd},\ref{eq:1/d}) and shown in Fig.\,\ref{fig:analytical_powerlaw_decay}.

\subsection{Dipole at the side of an ellipse}
\begin{figure}[htb]
	\begin{center}
		\includegraphics[width=1 \columnwidth]{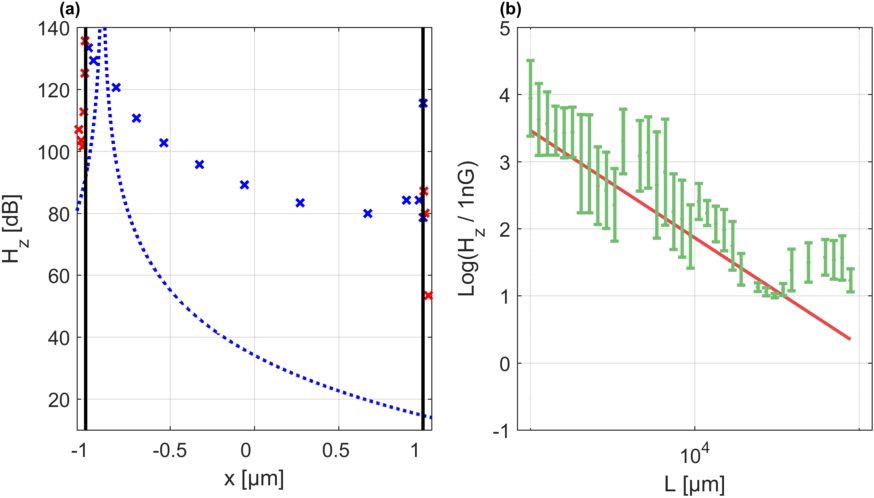}
		\protect\caption{(a) The effective magnetic field $H_z$ inside an elliptical aperture with $a = 1000 \nanometer$ and $b = 50 \nanometer$ caused by a dipole located at a distance of $d=100 \nanometer$ from the left edge of the aperture. The field is shown at the line $y = 5 \nanometer$. The y axis is the magnetic field, in dB, relative to 1 Gauss. The black lines indicate the edges of the aperture. The dashed curve shows the normal dipole decay. The points are the results of the simulation. A blue (red) point indicates that the field is in the negative (positive) $\hat{z}$ direction. (b) The effective magnetic field $H_z$ measured at a distance $d$ from the right edge of the aperture when the dipole is located at a distance $d$ from the left edge as function of the distance between the dipole and the measured point. The line is fitted to the log of the data and the slope is $-1.4 \pm 0.3$.}\label{fig:ellipse_simulation}
	\end{center}
\end{figure}

We proceed to simulate an elliptical aperture with the $x$ radius being $a = 1000 \nanometer$ and the $y$ radius $b = 100 \nanometer$. The dipole was located at a distance of $d=100 \nanometer$ from the left edge. The results are shown in Fig.\,\ref{fig:ellipse_simulation}(a). It can be seen that the behavior is similar to the round aperture, but that the magnetic field is stronger. We then expanded the simulations to elliptical apertures with a constant $b = 100 \nanometer$ and varying $a$. The magnetic field at a constant distance of $d$ from the right edge of the aperture was extracted from the data. The results [Fig.\,\ref{fig:ellipse_simulation}(b)] were fitted to a power-law scaling, obtaining a power law of $-1.4 \pm 0.3$.

\section{Discussion and Summary}
This work addresses the challenge of achieving strong magnetic coupling between magnetic dipoles, a major obstacle for scaling spin-based qubits, and an important aspect of non-local magnetic sensing.
To this end we derived an analytical solution for the flux confinement and for the magnetic field inside an aperture in a superconducting film. We have shown that for two dipoles next to the edges of a round aperture, the magnetic field, and thus the interaction, decays as $\frac{1}{r^2}$, which represents a significant improvement over the $\frac{1}{r^3}$ scaling of the normal dipole-dipole interaction. We have extended these results using numerical methods to show that the scaling improves even further for an elliptical aperture, reaching $\frac{1}{r^{3/2}}$. Other structures, such as the proposed ``dog-bone'' [Fig.\,\ref{fig:qubit_dogbone}] is expected to exhibit similar behavior as the ellipse due to the strong confinement in one of the axes.

A similar ``dog-bone'' structure, with a ferromagnet instead of a superconductor, has already been shown to improve the coupling by a ratio of $\frac{L}{D}$\cite{dogbone_ferromagnet_2013}, with $L$ being the distance between the qubits and $D$ being the radius of the ``dog-bone`` circular aperture. This leads to a power-law dependence of -2, which we have shown that can be achieved using a simpler circular aperture inside a superconducting film. By creating an ellipse or a ``dog-bone'' structure in a superconductor as proposed here, we expect to improve the power-law dependence even further to around -1.5. We also stress that the ferromagnetic coupler proposed in \cite{dogbone_ferromagnet_2013} introduces significant spin noise which can reduce the coherence time of the qubits, while the superconducting structure can avoid this adverse effect.

Two other works have shown direct NV-NV coupling between two NVs at close proximity. In one, two NVs at a distance of $10 \nanometer$ were shown to have a coupling of $\sim 40 \kilohertz$\cite{neumann_Quantum_register_2010}. In the other, two NVs at a distance of $22 \nanometer$ were shown to have a coupling of $\sim 4.9 \kilohertz$. Using the method proposed in this work, we can achieve a coupling of $\sim 10 \kilohertz$ at a distance of $300 \nanometer$. Such a coupling strength is significant compared to previous results mentioned above, and compared to NV coherence times achieved at cryogenic temperatures using dynamical decoupling schemes, reaching nearly 1 second \cite{Bar-Gill2013}. Thus, the proposed structure could allow high-fidelity two-qubit gates (at a rate of over 1000 operations within the coherence time), while maintaining a spatial separation between qubits of $\sim 300 \nanometer$, well beyond the optical diffraction limit, allowing straightforward addressing of individual NV centers.

These results could significantly impact several fields, including SQUIDs \cite{brandt_2005}, magnetic levitation \cite{Levitation}, quantum sensing and quantum computation \cite{romach_OSA_2017,semiconductor_review_2020}. 
Specifically, our findings could potentially allow coherent long range dipole-dipole interactions, which could pave the way toward scalable quantum computing in solid-state spin qubit architectures.

While the experimental realization of the proposed system is not trivial, our scheme can be scaled to larger magnetic dipole and aperture sizes, to facilitate simpler demonstrations. It is possible to experimentally verify our results by scaling all of the relevant sizes to $\micrometer$ or even $\millimeter$ scales. This could be done, for example, by placing a micrometer-sized ferromagnetic particle near one edge of a larger aperture, and measuring the local magnetic field near the opposite edge of the aperture, as a function of aperture size. These experiments, which are still not trivial and are currently being pursued, are expected to be presented in future publications.

We note that while our analysis is semi-classical, it is sufficient to prove that such long-range entanglement is possible. In accordance with the correspondence principle, it is impossible for the classical field to scale differently than the quantum interaction strength. 

We also note that in the context of quantum processing, an open question remains regrading the coherence of the enhanced dipole-dipole interaction mediated by the superconductor. Due to the phase preserving nature of the supercurrent, we expect that barring additional noise sources, the interaction will be coherent. Moreover, the superconducting gap for an $80 \nanometer$ thick Nb based superconductor is in the $300 \gigahertz$ range \footnote{The critical temperature for an $80 \nanometer$ Nb film is $\approx 9\,^{o}K$. The relation between the superconducting gap at zero temperature and the critical temperature, according to BCS theory, is $\Delta_0 = 1.75 k_B T_c$ \cite{bardeen_theory_1957}. The superconducting gap dependence on temperature is $\Delta(t) = \Delta_0 \text{tanh}\left( 1.75 \sqrt{\frac{T_c}{T}-1} \right)$ \cite{bardeen_theory_1957}. Plugging all of the above and assuming a temperature of $5\,^{o}K$, we get $\Delta = 300.7 \gigahertz$}, which is two orders of magnitude higher than the relevant energy scale for solid-state spin qubits (such as NV centers). This implies that excitations in the superconductor should not decohere the quantum spins. Other sources of noise arising from an imperfect superconducting layer will also affect the qubits. However, this noise will not be different from the noise affecting other superconducting circuits and should not change the scaling of the interaction. We therefore have strong indications that coherent coupling and entanglement will be realistically possible using this approach. In addition, improved fabrication techniques should be able to suppress these noise sources even further. Nevertheless, proof that coherent coupling is possible using this method will require further study and will be the subject of future work.

\section{Acknowledgment}
This work was originally motivated by discussions with Amir Yacoby.
N.B. acknowledges support from the European Union’s Horizon 2020 research and innovation program under grant agreements No. 714005 (ERC StG Q-DIM-SIM), No. 820374 (MetaboliQs), and No. 828946 (PATHOS), and has been supported in part by the Ministry of Science and Technology, Israel.  Y.R. is grateful for the support from the Kaye Einstein Scholarship and from the CAMBR fellowship.

\bibliography{NV}

\onecolumngrid
\appendix
\section{Appendix}

\subparagraph{Green function} \label{app:GreenFunction}
\hfill \break

Applying the Kelvin inversion transformation to the Green function of a semi-infinite film \cite{Semi_Infinite} results in the following Green function of an infinite film with a circular aperture (in cylindrical coordinates $(\rho,\phi,z)$:

{\footnotesize{}
\begin{equation*}
G(\boldsymbol{r},\boldsymbol{r}\boldsymbol{'})=\frac{1}{8\pi\epsilon_{0}}\left\{ \frac{1}{D_{-}}\left[1+\frac{2}{\pi}tan^{-1}\left(\frac{F_{-}}{D_{-}}\right)\right]-\frac{1}{D_{+}}\left[1+\epsilon\frac{2}{\pi}tan^{-1}\left(\frac{F_{+}}{D_{+}}\right)\right]\right\}
\end{equation*}
}
Where,
{\footnotesize{}
\[
F_{\pm}=\frac{1}{\sqrt{2}R}\left\{ \left(\rho^{2}+z^{2}-R^{2}\right)\left(\rho'^{2}+z'^{2}-R^{2}\right)\mp4R^{2}zz'+\sqrt{\left[z^{2}+\left(\rho+R\right)^{2}\right]\left[z^{2}+\left(\rho-R\right)^{2}\right]\left[z'^{2}+\left(\rho'+R\right)^{2}\right]\left[z'^{2}+\left(\rho'-R\right)^{2}\right]}\right\} ^{1/2}
\]
}{\footnotesize\par}

{\footnotesize{}
\[
D_{\pm}=\sqrt{\rho^{2}+\rho'^{2}-2\rho\rho'cos(\phi-\phi')+(z\pm z')^{2}},\,\,\,\,\,\epsilon=Sign\left[z\left(\rho'^{2}+z'^{2}-R^{2}\right)+z'\left(\rho{}^{2}+z{}^{2}-R^{2}\right)\right]
\]
}\\
With $R$ being the radius of the aperture.

\subparagraph{Calculating the magnetic field for a circular aperture with a centered dipole} \label{app:CalcField}
\hfill \break

Plugging in the expression for the current density $\boldsymbol{J}(\boldsymbol{r})\boldsymbol{=}-\boldsymbol{m}\times\nabla\delta(\boldsymbol{r})$ in Eq.\,\ref{eq:B_from_int_JG}:
\[
\boldsymbol{A}(\boldsymbol{r})=\mu_{0}\epsilon_{0}\int_{V}d^{3}r'\boldsymbol{J}(\boldsymbol{r}\boldsymbol{'})G(\boldsymbol{r},\boldsymbol{r}\boldsymbol{'})=-\mu_{0}\epsilon_{0}\boldsymbol{m}\times\int_{V}d^{3}r'\nabla_{\boldsymbol{r'}}\delta(\boldsymbol{r'})G(\boldsymbol{r},\boldsymbol{r}\boldsymbol{'})
\]
\[
=-\mu_{0}\epsilon_{0}\boldsymbol{m}\times\left[\int_{\sigma}da'\hat{n'}\delta(\boldsymbol{r'})G(\boldsymbol{r},\boldsymbol{r}\boldsymbol{'})-\int_{V}d^{3}r'\delta(\boldsymbol{r'})\nabla_{\boldsymbol{r'}}G(\boldsymbol{r},\boldsymbol{r}\boldsymbol{'})\right]
\]
The left term goes away due to $G(\boldsymbol{r},\boldsymbol{r}\boldsymbol{'})$ vanishing on $\sigma$, according to the boundary conditions. Now we have to calculate \mbox{$\nabla_{\boldsymbol{r'}}G(\boldsymbol{r},\boldsymbol{r}\boldsymbol{'})|_{\boldsymbol{r}\boldsymbol{'}=\boldsymbol{0}}$}.
From symmetry it is clear that we can't have a $\hat{\phi}$ component. Our approach will be assuming that $z,z'$ are positive, and then since the vector potential is continuous we can plug in $z'=0$. As we also know that it would be symmetrical for negative values, we will get the solution for all $z$. With these assumptions, $\epsilon|_{\boldsymbol{r}'=\boldsymbol{0}}\rightarrow-1$, and its $z'$ derivative will include a delta function that vanishes for $z'>0$. So from now on we can plug in $\epsilon=-1$. The derivative calculation in respect to variable $x'$ yields:
\[
8\pi\epsilon_{0}\partial_{x'}G=-\frac{\partial_{x'}D_{-}}{D_{-}^{2}}\left[1+\frac{2}{\pi}tan^{-1}\left(\frac{F_{-}}{D_{-}}\right)\right]+\frac{1}{D_{-}}\frac{2}{\pi}\frac{1}{1+\left(\frac{F_{-}}{D_{-}}\right)^{2}}\frac{\partial_{x'}F_{-}D_{-}-F_{-}\partial_{x'}D_{-}}{D_{-}^{2}}-
\]
\[
\left(-\frac{\partial_{x'}D_{+}}{D_{+}^{2}}\left[1-\frac{2}{\pi}tan^{-1}\left(\frac{F_{+}}{D_{+}}\right)\right]-\frac{1}{D_{+}}\frac{2}{\pi}\frac{1}{1+\left(\frac{F_{+}}{D_{+}}\right)^{2}}\frac{\partial_{x'}F_{+}D_{+}-F_{+}\partial_{x'}D_{+}}{D_{+}^{2}}\right)
\]
Now we will calculate $F_{\pm}|_{\boldsymbol{r}\boldsymbol{'}=\boldsymbol{0}}$,$D_{\pm}|_{\boldsymbol{r}\boldsymbol{'}=\boldsymbol{0}}$,
and their derivatives with respect to $\rho'$ and $z'$:
\[
F_{\pm}|_{\boldsymbol{r}\boldsymbol{'}=\boldsymbol{0}}\equiv F=\frac{1}{\sqrt{2}}\left\{ -\left(\rho^{2}+z^{2}-R^{2}\right)+\sqrt{\left[z^{2}+\left(\rho+R\right)^{2}\right]\left[z^{2}+\left(\rho-R\right)^{2}\right]}\right\} ^{1/2}
\]
\[
D_{\pm}|_{\boldsymbol{r}\boldsymbol{'}=\boldsymbol{0}}\equiv D=\sqrt{\rho^{2}+z{}^{2}}=r
\]
\[
\partial_{\rho'}F_{\pm}|_{\boldsymbol{r}\boldsymbol{'}=\boldsymbol{0}}=0,\,\,\,\,\,\partial_{z'}F_{z}|_{\boldsymbol{r}\boldsymbol{'}=\boldsymbol{0}}=\frac{\mp Rz}{\sqrt{2}F},\,\,\,\,\,\partial_{\rho'}D_{\pm}|_{\boldsymbol{r}\boldsymbol{'}=\boldsymbol{0}}=-\frac{\rho}{D},\,\,\,\,\,\partial_{z'}D_{\pm}|_{\boldsymbol{r}\boldsymbol{'}=\boldsymbol{0}}=\frac{\pm z}{D}
\]
We choose $\phi'=\phi$, which will give us later $\hat{\rho}'=\hat{\rho}$.
Plugging in $F_{\pm}|_{\boldsymbol{r}\boldsymbol{'}=\boldsymbol{0}},D_{\pm}|_{\boldsymbol{r}\boldsymbol{'}=\boldsymbol{0}}$ we get:
\begin{multline*}
8\pi\epsilon_{0}x'G|_{\boldsymbol{r}\boldsymbol{'}=\boldsymbol{0}}=\\
\frac{1}{D^{2}}\left(-\partial_{x'}D_{-}\left[1+\frac{2}{\pi}tan^{-1}\left(\frac{F}{D}\right)\right]+\partial_{x'}D_{+}\left[1-\frac{2}{\pi}tan^{-1}\left(\frac{F}{D}\right)\right]\right)+\frac{2\left(\left(\partial_{x'}F_{+}+\partial_{x'}F_{-}\right)D-\left(\partial_{x'}D_{+}+\partial_{x'}D_{-}\right)F\right)}{\pi D^{3}\left(1+\left(\frac{F}{D}\right)^{2}\right)}
\end{multline*}
And now writing the full gradient, and plugging in the derivatives of the terms:
\[
\nabla_{\boldsymbol{r'}}G(\boldsymbol{r},\boldsymbol{r}\boldsymbol{'})|_{\boldsymbol{r}\boldsymbol{'}=\boldsymbol{0}}=\partial_{\rho'}G|_{\boldsymbol{r}\boldsymbol{'}=\boldsymbol{0}}\hat{\rho}+\partial_{z'}G|_{\boldsymbol{r}\boldsymbol{'}=\boldsymbol{0}}\hat{z}=\frac{1}{4\pi r^{3}}\left[\frac{2}{\pi}\left(tan^{-1}\left(\alpha(\boldsymbol{r})\right)+\frac{\alpha(\boldsymbol{r})}{1+\alpha(\boldsymbol{r})^{2}}\right)\rho\hat{\rho}+z\hat{z}\right]\equiv\boldsymbol{n}(\boldsymbol{r})
\]
Where,
\[
\alpha(\boldsymbol{r})\equiv\frac{F}{D}=\frac{1}{\sqrt{2}r}\sqrt{R^{2}-r^{2}+\sqrt{\left[z^{2}+\left(\rho+R\right)^{2}\right]\left[z^{2}+\left(\rho-R\right)^{2}\right]}}
\]
\\
Plugging back we now get:
\begin{equation*}
\boldsymbol{A}(\boldsymbol{r})=\mu_{0}\epsilon_{0}\boldsymbol{m}\times\nabla_{\boldsymbol{r'}}G(\boldsymbol{r},\boldsymbol{r}\boldsymbol{'})|_{\boldsymbol{r}\boldsymbol{'}=\boldsymbol{0}}=\frac{\mu_{0}}{4\pi}\frac{\boldsymbol{m}\times\boldsymbol{\hat{n}}(\boldsymbol{r})}{r^{2}}
\end{equation*}
Now for the magnetic field, taking curl we get:
\[
\boldsymbol{B}(\boldsymbol{r})=\frac{\mu_{0}}{4\pi}\nabla\times\left(\frac{\boldsymbol{m}\times\boldsymbol{n}(\boldsymbol{r})}{r^{3}}\right)=\frac{\mu_{0}}{4\pi}\left[\frac{1}{r^{3}}\nabla\times\left(\boldsymbol{m}\times\boldsymbol{n}(\boldsymbol{r})\right)-\frac{3}{r^{5}}\boldsymbol{r}\times\left(\boldsymbol{m}\times\boldsymbol{n}(\boldsymbol{r})\right)\right]
\]
\[
=\frac{\mu_{0}}{4\pi}\left[\frac{1}{r^{3}}\left(\left(\nabla\cdot\boldsymbol{n}(\boldsymbol{r})\right)\boldsymbol{m}-\left(\boldsymbol{m}\cdot\nabla\right)\boldsymbol{n}(\boldsymbol{r})\right)-\frac{3}{r^{5}}\left(\text{\ensuremath{\left(\boldsymbol{r}\cdot\boldsymbol{n}(\boldsymbol{r})\right)}}\boldsymbol{m}-\left(\boldsymbol{m}\cdot\boldsymbol{r}\right)\boldsymbol{n}(\boldsymbol{r})\right)\right]
\]
\begin{equation*}
=\frac{\mu_{0}}{4\pi r^{3}}\left[3\left(\boldsymbol{m}\cdot\boldsymbol{\hat{r}}\right)\hat{\boldsymbol{n}}(\boldsymbol{r})-\left(\boldsymbol{m}\cdot\nabla\right)\boldsymbol{n}(\boldsymbol{r})+\left(\nabla\cdot\boldsymbol{n}(\boldsymbol{r})-3\left(\hat{\boldsymbol{r}}\cdot\boldsymbol{\hat{n}}(\boldsymbol{r})\right)\right)\boldsymbol{m}\right]
\end{equation*}

\begin{figure}[htb]
(a) $\boldsymbol{m}=m\hat{z}$~~~~~~~~~~~~~~~~~~~~~~~~~~~~~~~~~~~~~~~~~~~~~~~~~~~~~~~~~~~~~~~~~~~(b) $\boldsymbol{m}=m\hat{z}$~~~$z=10^{-3}R$\\ \includegraphics[scale=0.8]{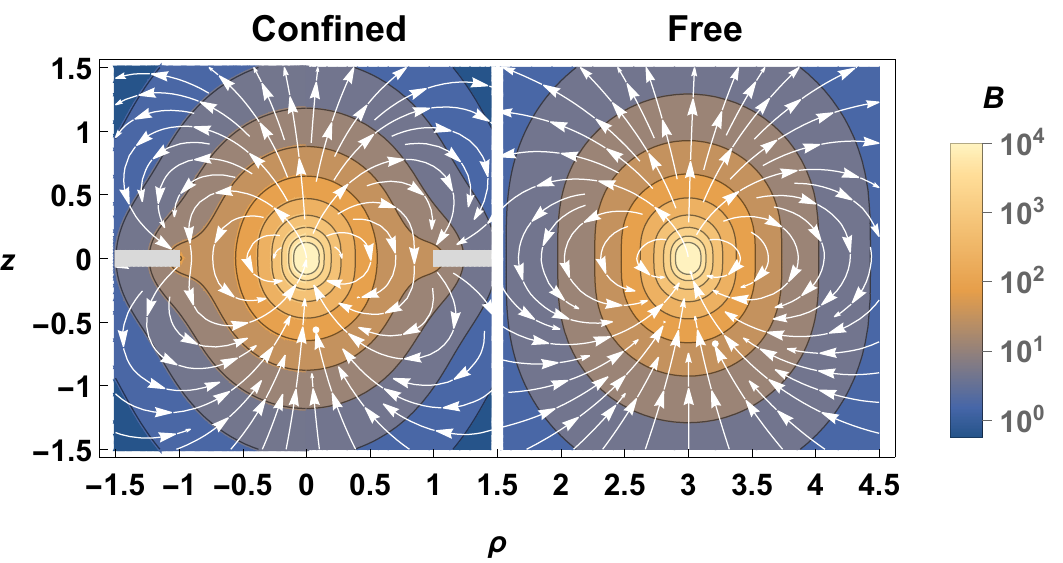}~~~~~~~\includegraphics[scale=0.8]{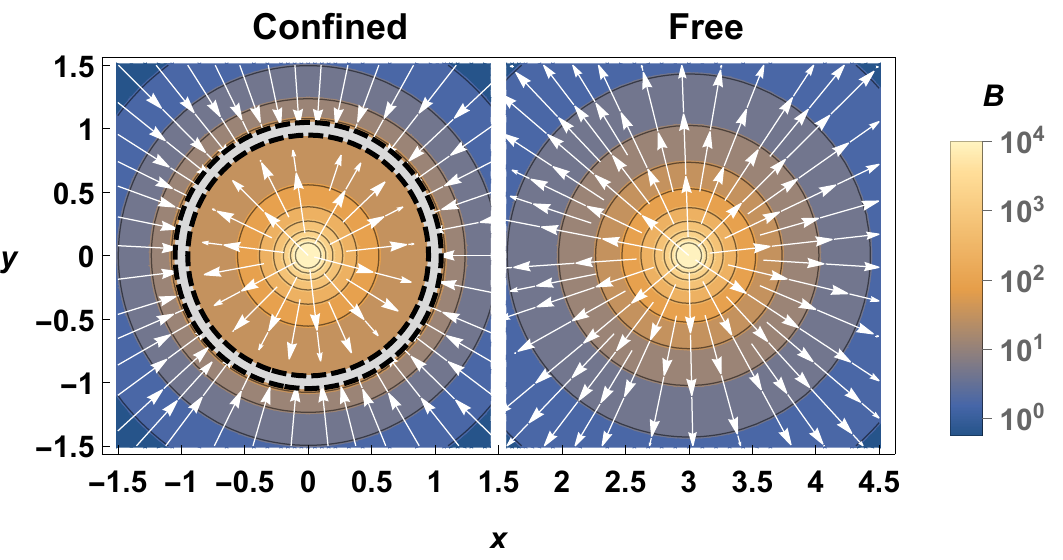}\\
(c) $\boldsymbol{m}=m\hat{z}$~~~zoom of (a)~~~~~~~~~~~~~~~~~~~~~~~~~~~~~~~~~~~~~~~~~~~~~~~~~(d) $\boldsymbol{m}=m\hat{x}$\\ \includegraphics[scale=0.8]{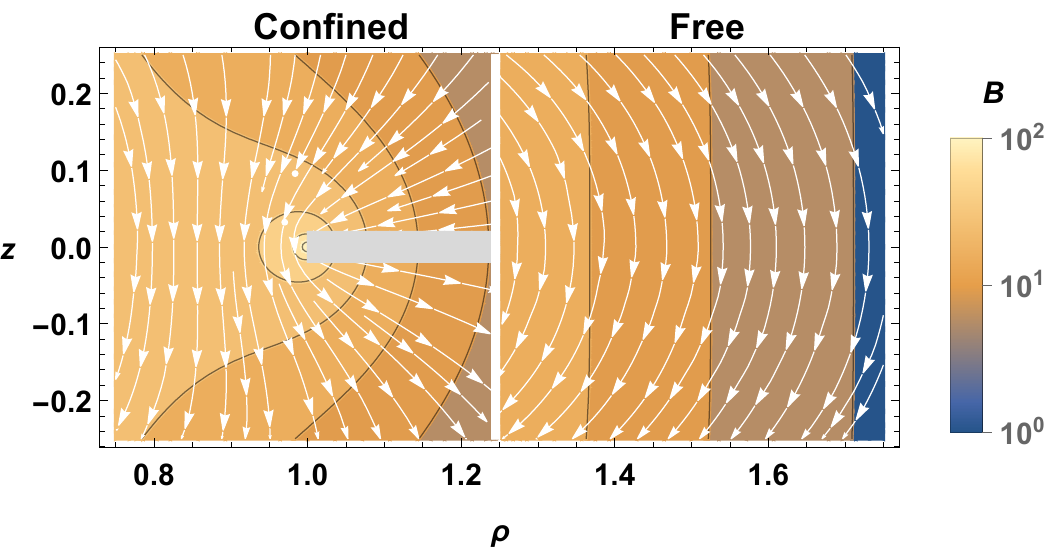}~~~~~~~\includegraphics[scale=0.8]{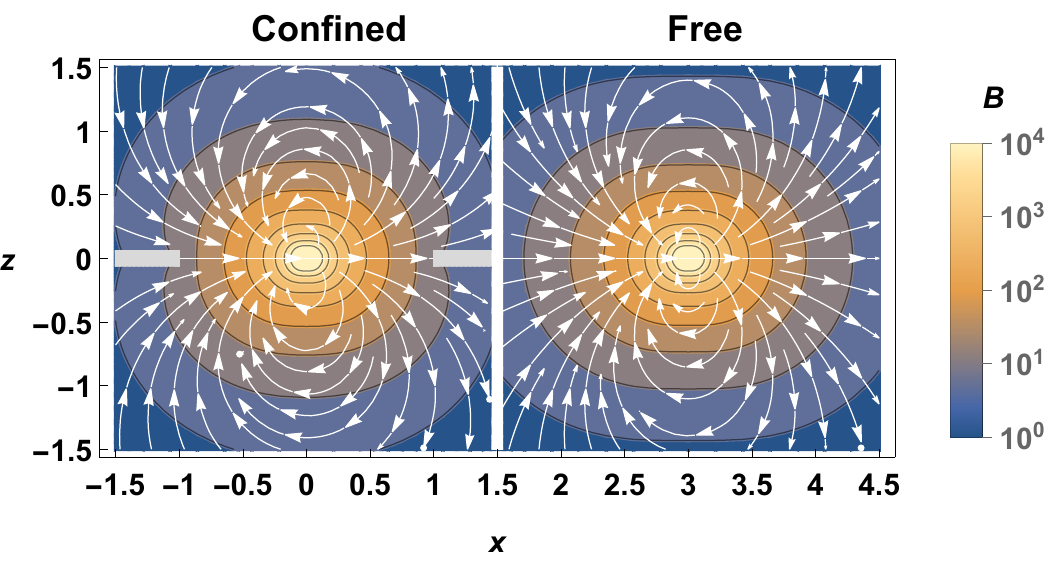}\\
(e) $\boldsymbol{m}=m\hat{x}$~~~$x=10^{-3}R$~~~~~~~~~~~~~~~~~~~~~~~~~~~~~~~~~~~~~~~~~~~~~~~~~~(f) $\boldsymbol{m}=m\hat{x}$~~~$z=10^{-3}R$\\
\includegraphics[scale=0.8]{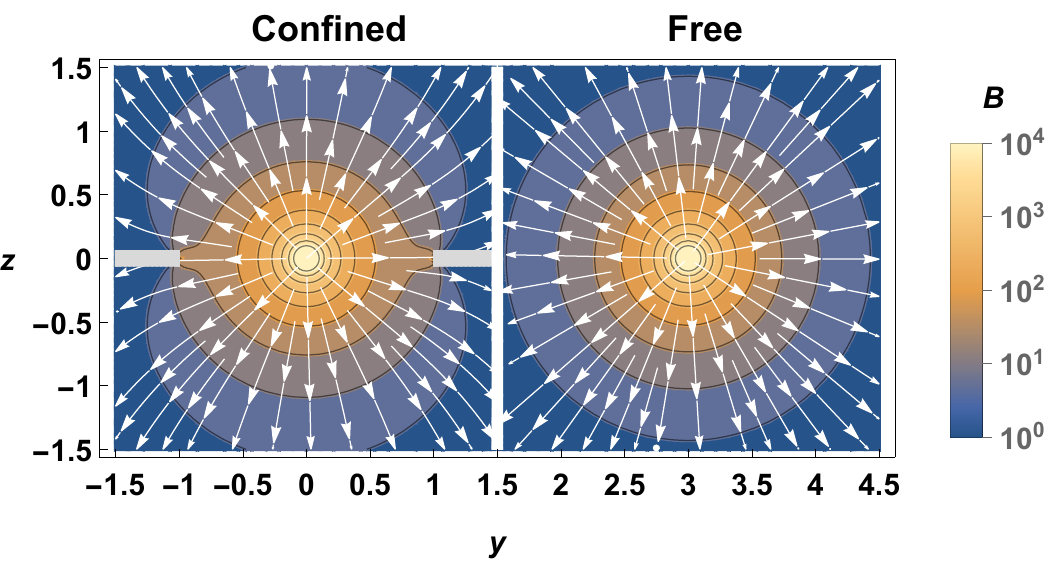}~~~~~~~\includegraphics[scale=0.8]{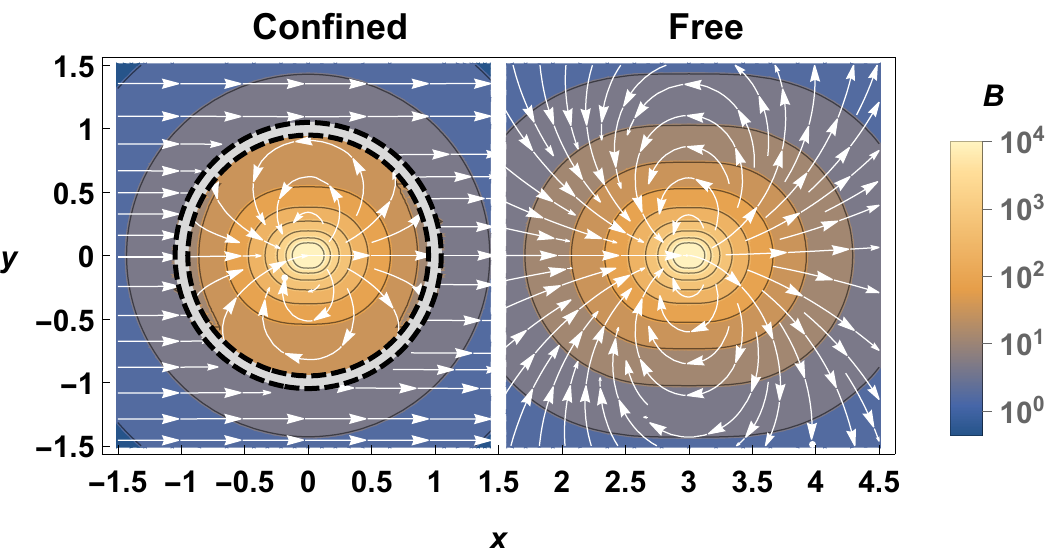}

\caption{The analytical solution for the magnetic field of a dipole in a circular aperture of a superconducting film, calculated from Eq.\,\ref{eq:B_for_centered_dipole}, for $\mu_{0}m=R=1$. The dipole in (a)-(c) is pointing towards $\hat{z}$, and in (d)-(f) is pointing towards $\hat{x}$. The stream plots provide the direction of the field, and the contour plots are scaled relative to the magnitude of the field. We see the comparison to the free magnetic dipole field in the right side plots. It is evident visually that the superconductor confines the magnetic flux. The flux confinement exhibits different geometric behaviors in different dipole orientations.}
\label{appfig:analytical_streamplot_for_round_hole}
\end{figure}

\subparagraph{Calculating \mbox{$\boldsymbol{B}_{\boldsymbol{m}}(z=0)$}} \label{app:FieldonPlane}
\hfill \break
\[
\boldsymbol{B}({z=}0)=\frac{\mu_{0}}{4\pi\rho^{3}}\left[3\left(\boldsymbol{m}\cdot\hat{\rho}\right)\hat{\boldsymbol{n}}_{\rho}(\rho)-\left(\boldsymbol{m}\cdot\nabla\right)\boldsymbol{n}(\boldsymbol{r})|_{z=0}+\left(\nabla\cdot\boldsymbol{n}(\boldsymbol{r})|_{z=0}-3\hat{n}_{\rho}(\rho)\right)\boldsymbol{m}\right]
\]
Assuming $\boldsymbol{m}=m_{x}\hat{x}+m_{z}\hat{z}$ (without loss of generality since the geometry $\sigma$ has an azimuthal symmetry):

{\footnotesize{}
\[
\boldsymbol{B}({z=}0)=\frac{\mu_{0}}{4\pi\rho^{3}}\left[3m_{x}cos\phi\frac{n_{\rho}(\rho)}{\rho}\hat{\rho}-\left(m_{x}\partial_{x}n_{x}(\rho)\hat{x}+m_{x}\partial_{x}n_{y}(\rho)\hat{y}+m_{z}\partial_{z}n_{\rho}(\boldsymbol{r})|_{z=0}\hat{\rho}+m_{z}\hat{z}\right)+\left(\frac{1}{\rho}\partial_{\rho}\left(\rho n_{\rho}(\rho)\right)+1-3\frac{n_{\rho}(\rho)}{\rho}\right)\left(m_{x}\hat{x}+m_{z}\hat{z}\right)\right]
\]
}{\footnotesize\par}

\[
\partial_{z}\alpha(\boldsymbol{r})|_{z=0}=0\Rightarrow\partial_{z}n_{\rho}(\boldsymbol{r})|_{z=0}=0
\]

\begin{multline*}
\boldsymbol{B}(z=0)=\frac{\mu_{0}}{4\pi\rho^{3}}\left[3m_{x}cos\phi\frac{n_{\rho}(\rho)}{\rho}\hat{\rho}-\left(m_{x}\partial_{x}\left(\frac{n_{\rho}(\rho)}{\rho}x\right)\hat{x}+m_{x}\partial_{x}\left(\frac{n_{\rho}(\rho)}{\rho}y\right)\hat{y}+m_{z}\hat{z}\right)\right.+\\
\left.\left(\partial_{\rho}n_{\rho}(\rho)+1-2\frac{n_{\rho}(\rho)}{\rho}\right)\left(m_{x}\hat{x}+m_{z}\hat{z}\right)\right]
\end{multline*}
\[
=\frac{\mu_{0}}{4\pi\rho^{3}}\left[m_{x}\left(3cos\phi\frac{n_{\rho}(\rho)}{\rho}\hat{\rho}-\partial_{x}\left(\frac{n_{\rho}(\rho)}{\rho}\right)\rho\hat{\rho}-\frac{n_{\rho}(\rho)}{\rho}\hat{x}\right)+m_{x}\hat{x}+\left(\partial_{\rho}n_{\rho}(\rho)-2\frac{n_{\rho}(\rho)}{\rho}\right)\left(m_{x}\hat{x}+m_{z}\hat{z}\right)\right]
\]
\[
\partial_{x}=\partial_{x}\rho\cdot\partial_{\rho}=cos\phi\partial_{\rho}
\]
\[
=\frac{\mu_{0}}{4\pi\rho^{3}}\left[m_{x}cos\phi\left(3\frac{n_{\rho}(\rho)}{\rho}-\partial_{\rho}\left(\frac{n_{\rho}(\rho)}{\rho}\right)\rho\right)\hat{\rho}+m_{x}\left(1-\frac{n_{\rho}(\rho)}{\rho}\right)\hat{x}+\left(\partial_{\rho}n_{\rho}(\rho)-2\frac{n_{\rho}(\rho)}{\rho}\right)\left(m_{x}\hat{x}+m_{z}\hat{z}\right)\right]
\]
\[
=\frac{\mu_{0}}{4\pi\rho^{3}}\left[m_{x}cos\phi\left(4\frac{n_{\rho}(\rho)}{\rho}-\partial_{\rho}n_{\rho}(\rho)\right)\hat{\rho}+m_{x}\left(1-\frac{n_{\rho}(\rho)}{\rho}\right)\hat{x}+\left(\partial_{\rho}n_{\rho}(\rho)-2\frac{n_{\rho}(\rho)}{\rho}\right)\left(m_{x}\hat{x}+m_{z}\hat{z}\right)\right]
\]
Calculating the different terms:

\[
\alpha(\rho,z=0)=\begin{cases}
\sqrt{\left(\frac{R}{\rho}\right)^{2}-1} & \rho<R\\
0 & R<\rho
\end{cases}\Rightarrow n_{\rho}(\rho)=\begin{cases}
\frac{2}{\pi}\left(cos^{-1}\left(\frac{\rho}{R}\right)+\frac{\rho}{R}\sqrt{1-\left(\frac{\rho}{R}\right)^{2}}\right)\rho & \rho<R\\
0 & R<\rho
\end{cases}
\]

\[
\partial_{\rho}n_{\rho}(\rho<R)=\frac{2}{\pi}\left(cos^{-1}\left(\frac{\rho}{R}\right)-\frac{\rho}{R\sqrt{1-\left(\frac{\rho}{R}\right)^{2}}}+\frac{2\rho}{R}\sqrt{1-\left(\frac{\rho}{R}\right)^{2}}+\frac{\rho^{2}}{R}\frac{-\frac{2\rho}{R^{2}}}{2\sqrt{1-\left(\frac{\rho}{R}\right)^{2}}}\right)
\]
\[
=\frac{2}{\pi}\left(cos^{-1}\left(\frac{\rho}{R}\right)-\frac{\frac{\rho}{R}\left(1+\left(\frac{\rho}{R}\right)^{2}\right)}{\sqrt{1-\left(\frac{\rho}{R}\right)^{2}}}+2\frac{\rho}{R}\sqrt{1-\left(\frac{\rho}{R}\right)^{2}}\right)
\]
\[
4\frac{n_{\rho}(\rho)}{\rho}-\partial_{\rho}n_{\rho}(\rho)=\frac{2}{\pi}\left(3cos^{-1}\left(\frac{\rho}{R}\right)+2\frac{\rho}{R}\sqrt{1-\left(\frac{\rho}{R}\right)^{2}}+\frac{\frac{\rho}{R}\left(1+\left(\frac{\rho}{R}\right)^{2}\right)}{\sqrt{1-\left(\frac{\rho}{R}\right)^{2}}}\right)
\]
\[
\partial_{\rho}n_{\rho}(\rho)-2\frac{n_{\rho}(\rho)}{\rho}=-\frac{2}{\pi}\left(cos^{-1}\left(\frac{\rho}{R}\right)+\frac{\frac{\rho}{R}\left(1+\left(\frac{\rho}{R}\right)^{2}\right)}{\sqrt{1-\left(\frac{\rho}{R}\right)^{2}}}\right)
\]
\[
\boldsymbol{B}(z=0)=\frac{\mu_{0}}{4\pi\rho^{3}}\begin{cases}
m_{x}cos\phi\left(4\frac{n_{\rho}(\rho)}{\rho}-\partial_{\rho}n_{\rho}(\rho)\right)\hat{\rho}+m_{x}\left(1-\frac{n_{\rho}(\rho)}{\rho}\right)\hat{x}+\left(\partial_{\rho}n_{\rho}(\rho)-2\frac{n_{\rho}(\rho)}{\rho}\right)\left(m_{x}\hat{x}+m_{z}\hat{z}\right) & \rho<R\\
m_{x}\hat{x} & R<\rho
\end{cases}
\]
Plugging in back the terms calculated in this expression and choosing orientation for $\boldsymbol{m}$ and angle $\phi$, this reduces to:

\begin{align}
\begin{split}
\boldsymbol{B}^{m\hat{z}}(\rho,\phi,0)=\frac{\mu_{0}m}{4\pi\rho^{3}} \hat{z} \begin{cases}
-\frac{2}{\pi}\left(cos^{-1}\left(\frac{\rho}{R}\right)+\frac{\frac{\rho}{R}\left(1+\left(\frac{\rho}{R}\right)^{2}\right)}{\sqrt{1-\left(\frac{\rho}{R}\right)^{2}}}\right) & \rho<R\\
0 & \rho>R
\end{cases} \\
\\
\boldsymbol{B}^{m\hat{y}}(x,0,0)=\frac{\mu_{0}m}{4\pi x^{3}} \hat{y} \begin{cases}
-\frac{2}{\pi}\left(2cos^{-1}\left(\frac{x}{R}\right) + \frac{2\frac{x}{R}}{\sqrt{1-\left(\frac{x}{R}\right)^{2}}}-\frac{\pi}{2}\right) & x<R\\
1 & x>R
\end{cases} \\
 \\
\boldsymbol{B}^{m\hat{x}}(x,0,0)=\frac{\mu_{0}m}{4\pi x^{3}} \hat{x} \begin{cases}
\frac{2}{\pi}\left(cos^{-1}\left(\frac{x}{R}\right)+\frac{x}{R}\sqrt{1-\left(\frac{x}{R}\right)^{2}}+\frac{\pi}{2}\right) & x<R\\
1 & x>R
\end{cases} 
\end{split}
\label{appeq:Bm_for_all_directionsx}
\end{align}

\begin{figure}[htb]
\begin{raggedright}
~~~~~~~~~~~~~~~(a) ~~~~~~~~~~~~~~~~~~~~~~~~~~~~~~~~~~~~~~~~~~~~~~~~~~~~~~~(b) \\
~~~~~~~~~~~~~~~\includegraphics[scale=0.65]{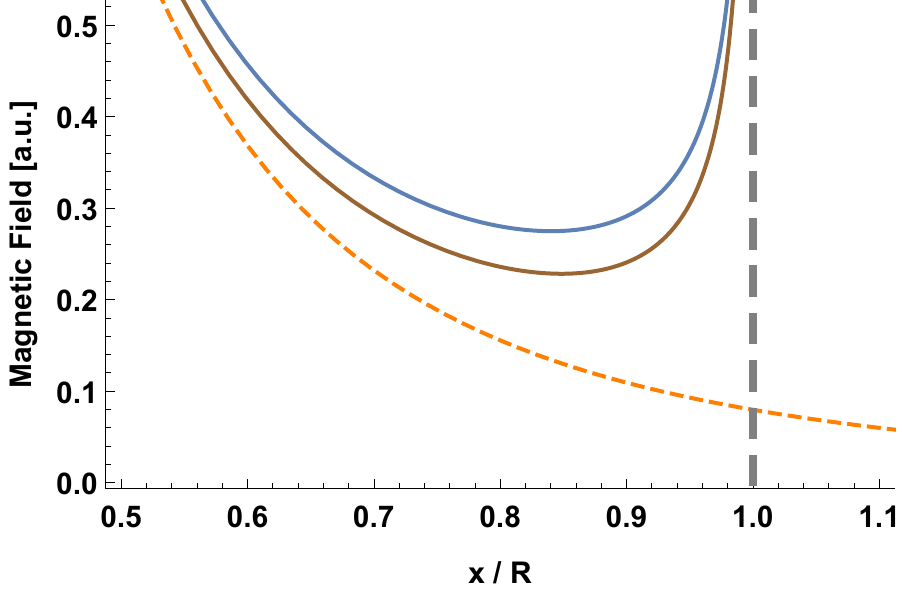}~~~~~~~\includegraphics[scale=0.65]{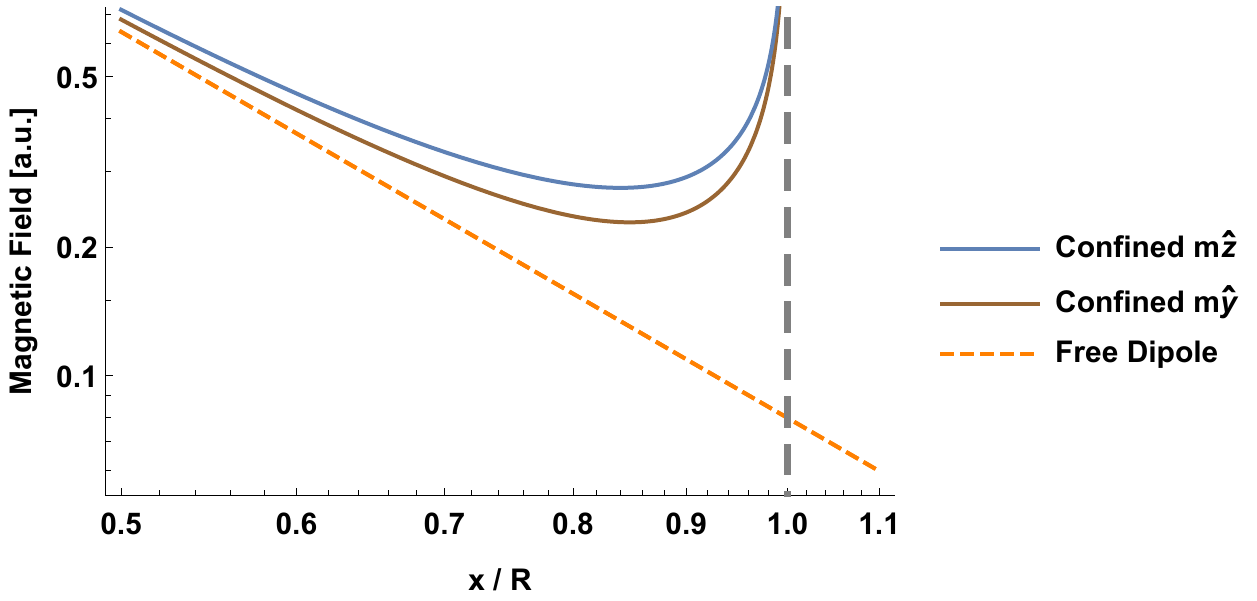}\\
~~~~~~~~~~~~~~~(c) ~~~~~~~~~~~~~~~~~~~~~~~~~~~~~~~~~~~~~~~~~~~~~~~~~~~~~~~(d) \\
~~~~~~~~~~~~~~~\includegraphics[scale=0.65]{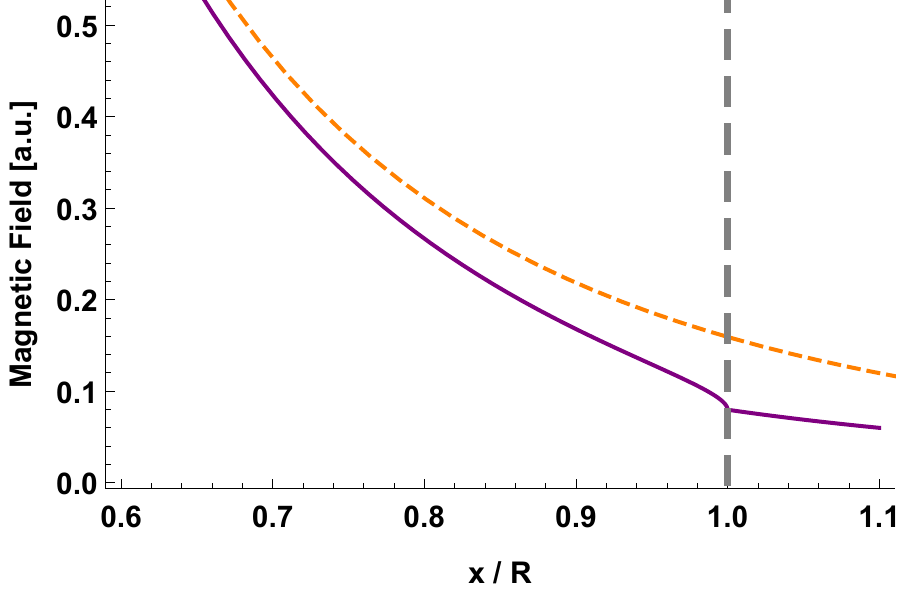}~~~~~~~\includegraphics[scale=0.65]{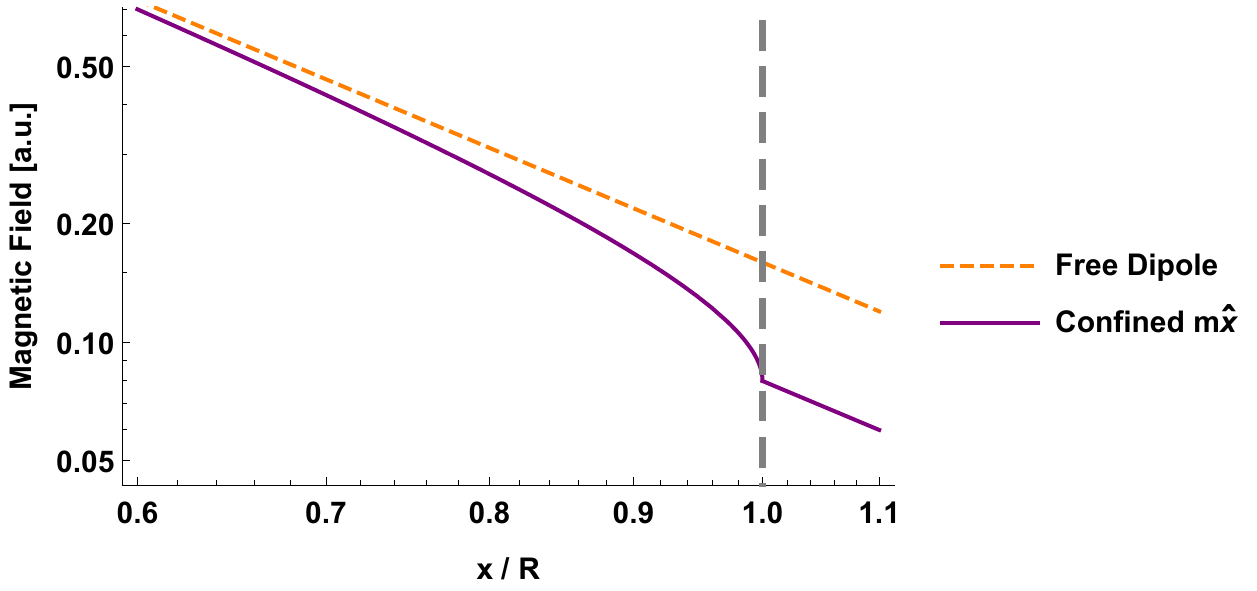}
\par\end{raggedright}
\caption{The analytical solution of the magnetic dipole field in a circular aperture of a superconducting film, calculated from Eq.\,\ref{appeq:Bm_for_all_directionsx}, with $\mu_{0}m=R=1$. (a,b) The magnetic field along the $x$ axis when the dipole is in $\hat{z}$ or $\hat{y}$ directions in (a) normal scale and (b) log scale. (c,d) The magnetic field along the $x$ axis when the dipole is in the $\hat{x}$ direction in (c) normal scale and (d) log scale. Note that the dipole field is a factor of 2 higher along it's axis. When the dipole is in the $\hat{z}$ direction, the field vanishes on the superconductor, since no field lines can cross it. When the dipole is in the $\hat{x}$ or $\hat{y}$ directions, the field does not vanish since the superconductor has zero thickness and can't block the field lines. For the $\hat{y}$ case, the field at $x>R$ coincide with the free dipole field and is not visible in the figure. Taking the limit of $R\rightarrow\infty$ (or $\rho\rightarrow0$), we get the behavior of a free dipole, with Eq.\,\ref{appeq:Bm_for_all_directionsx} approaching the free dipole case. We see that when the dipole is in the $\hat{z}$ or $\hat{y}$ directions the magnetic field behave quiet similar, approaching infinity as we get closer to the edge, with the dipole in the $\hat{z}$ resulting in a  bit stronger magnetic field. A dipole in the $\hat{x}$ direction results in a weaker field compared to the free one.}
\label{appfig:analytical_B_for_round_hole}
\end{figure}

\subparagraph{Numerical simulations}
\label{app:Simulations}
\hfill \break
Ampere's law for the field around a superconducting film can be written as:
\begin{equation}
H_z (\boldsymbol{r}) = H_a (\boldsymbol{r}) + \int_S{d^2 r'} Q(\boldsymbol{r}, \boldsymbol{r'}) g(\boldsymbol{r'}))
\label{appeq:Ampere_law}
\end{equation}
With $H_z (\boldsymbol{r})$ being the resulting perpendicular magnetic field, $H_a (\boldsymbol{r})$ being the applied perpendicular magnetic field, $g(\boldsymbol{r'})$ is the current stream function and  $Q(\boldsymbol{r}, \boldsymbol{r'})$ is a kernel that represents the perpendicular magnetic field created at point $\boldsymbol{r}$ by a magnetic dipole of unit strength and $\hat{z}$ direction at position $\boldsymbol{r'}$ .
The second London equation in 2D:
\begin{equation}
H_z (x,y) = - \Lambda \bigl[\nabla \times \boldsymbol{J}(x,y)\bigr]\hat{z} = \Lambda \nabla^2 g(x,y)
\label{appeq:2d_London}    
\end{equation}
Eliminating $Hz$ from Eq.\,\ref{appeq:2d_London} using Eq.\,\ref{appeq:Ampere_law} we can get (in discretized form):

\begin{equation}
H_a (\boldsymbol{r_i}) = - \sum_j{\bigl( Q_{ij}w_j - \Lambda \nabla^2_{ij} \bigr) g(\boldsymbol{r_j})}
\label{appeq:HaFromG}    
\end{equation}
$w_j$ is the weight of point $(\boldsymbol{r_j})$ which represent its 2d volume.
The matrix $\nabla^2_{ij}$ computes the 2D Laplacian at point $\boldsymbol{r_i}$ from the stream function $g(\boldsymbol{r_j})$ and its four nearest neighbors.
By inverting this equation, we can directly extract the stream function from the applied magnetic field:
\begin{equation}
G (\boldsymbol{r_i}) = - \sum_j{K_{ij} H_a (\boldsymbol{r_j})}
\label{appeq:GFromHa}    
\end{equation}
With the inverse matrix $K_{ij}$:
\begin{equation}
K_{ij} = \bigl( Q_{ij}w_j - \Lambda \nabla^2_{ij} \bigr)^{-1}
\label{appeq:InverseK}    
\end{equation}
This matrix inversion is the most resource intensive process of the simulation and is the limiting factor of the grid size.
The simulation works by first calculating the stream function $g$ using Eq.\,\ref{appeq:GFromHa} and then extracting the resulting magnetic field using Eq.\,\ref{appeq:Ampere_law}.

The stream function $g(x,y)$ has several useful properties:
\begin{enumerate}
    \item The current in the superconductor flows on the contour lines of $g(x,y)$.
    \item $g(x,y)$ is constant outside the film and inside apertures.
    \item As $g(x,y)$ is a scalar potential, it is defined up to a constant. We choose $g(x,y) = 0$ outside the superconducting film.
    \item With the previous choice of the constant, we find inside an isolated aperture $g(x,y) = I_0$, where $I_0$ is the current that circulates the aperture.
\end{enumerate}
For a more detailed explanation, see \cite{brandt_2005}.

\begin{figure}[htb]
	\begin{center}
		\includegraphics[width=1 \columnwidth]{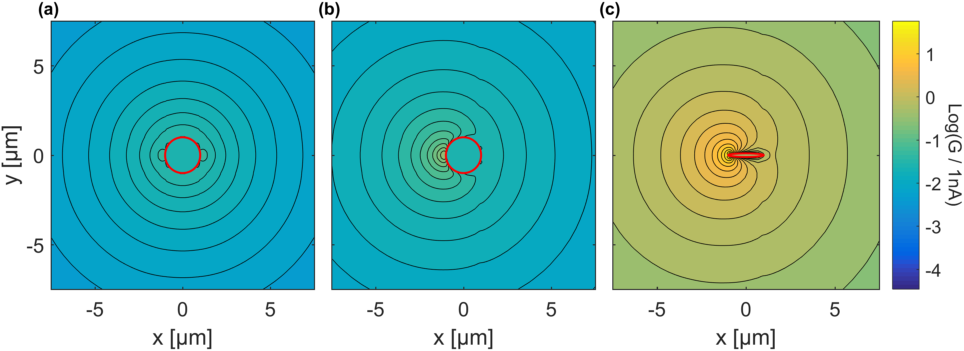}
		\protect\caption{The current stream function $G(x,y)$, in log scale, as calculated by the simulation for the cases in the main text: (a) a dipole in the middle of a round aperture [Fig.\,\ref{fig:simulation_analytic}(a)], (b) a dipole in the side of a round aperture [Fig.\,\ref{fig:simulation_analytic}(b)], (c) a dipole in the side of an elliptical aperture [Fig.\,\ref{fig:ellipse_simulation}(a)].
		The red line shows the edges of the aperture.
		} \label{appfig:numerical_streamplots}
	\end{center}
\end{figure}

The simulated stream function $g(x,y)$ for the different scenarios discussed in this article can be seen in Fig.\,\ref{appfig:numerical_streamplots}.

\end{document}